\documentclass[draft]{agujournal2019}
\usepackage{url} 
\usepackage{lineno}
\usepackage[inline]{trackchanges} 
\usepackage{soul}
\usepackage[export]{adjustbox}
\usepackage{amsmath}
\draftfalse

\journalname{arXiv}

\begin{document}

\title{Training Image Selection using Recurrent Neural Networks: An Application in Hydrogeology}

%
%




\authors{Zhendan Cao\affil{1}, Jiawei Shen\affil{2}\thanks{This author shares the same contribution with the first author}, Mathieu Gravey\affil{3}}


\affiliation{1}{Department of Land, Air and Water Resources. University of California, Davis, USA}
\affiliation{2}{College of Mathematics and Statistics. Chongqing University, Chongqing, China}
\affiliation{3}{Institute of Earth Surface Dynamics, University of Lausanne, Lausanne, Switzerland}




\correspondingauthor{Zhendan Cao}{zdcao@ucdavis.edu}




\begin{keypoints}
\item A new training image(TI) selection model for multiple point geostatistics is proposed based on recurrent neural networks(RNNs)
\item RNNs are applied to extract features from the observed hydraulic head time series as the criteria for the TI selection
\item The GRU model has the best performance in the TI selection task and can reach to a 97.63\% accuracy
\end{keypoints}

%
%

%
%


\begin{abstract}

Multiple-point geostatistics plays an important role in characterizing complex subsurface aquifer systems such as channelized structures. However, only a few studies have paid attention to how to choose an applicable training image. In this paper, a TI selection method based on Recurrent Neural Networks is proposed. A synthetic case is tested using two channelized training images given the hydraulic head time series. Three different RNNs architectures are tested for the selection performance. Various scenarios of the model input are also tested including the number of observation wells, the observation time steps, the influence of observation noise and the training dataset size. In this TI selection task, the GRU has the best performance among all three architectures and can reach to a $97.63\%$ accuracy.
\end{abstract}

\section*{Plain Language Summary}
The physical properties of a subsurface aquifer system such as hydraulic conductivity plays a significant role in groundwater modeling. Multiple point geostatistics(MPS) is a competent approach for reproducing complex subsurface systems. A Training Image(TI) can be a conceptual representation of the subsurface hydraulic conductivity and it is the foremost input in a MPS algorithm. However, few studies have paid attention to how to choose a proper TI. In this paper, we proposed a new method using observations of the hydraulic head to select the most likely TI from the TI candidates with the help of Recurrent Neural Networks. Two different channelized training images and three different RNNs architectures were applied to test the performance of this new method. The result shows that our method can have a 97.63\% accuracy in the TI selection task.

\section{Introduction}
Geological subsurface modeling plays a significant role in groundwater protection and management. Reproducing the strong heterogeneity of aquifers is one of the key tasks of most studies. Geostatistical simulations can capture the details of the heterogeneity of subsurface models. However, the traditional two-point based geostatistical modeling failed in reproducing complex subsurface structures such as channelized geometries\cite{caers2001geostatistical,journel2006necessity}. 

Multiple point geostatistics (MPS) has gained its popularity for capturing the complexity of the subsurface systems. The widely-used algorithms are SNESIM \cite{strebelle2002conditional}, Direct Sampling\cite{mariethoz2010direct}, and Graph Cuts methods\cite{li2016patch}. The main idea of those MPS algorithms is generating realizations which can mimic the Training Image(TI). The TI image is a conceptual model which can reflect the geometrical structure of the subsurface and it can be obtained from outcrop data, object-based methods \cite{deutsch1996hierarchical,holden1998modeling} and process-based methods\cite{Gross1998process,PYRCZ20091671}.

However, as the foremost part in the MPS workflow, TI is usually selected by researchers from their empirical understanding of the study area based on the outcrops and sparse hard data ,and therefore, often with a huge uncertainty which can cause a great impact to the following studies\cite{tahmasebi2018multiple}. Nevertheless, only a few studies have discussed how to choose or determine a proper TI. \citeA{brunetti2019hydrogeological} proposed a MCMC based method to select the hydrogeological model and have an application on the MADE site. However, due the computational cost, it is not feasible to apply this method to a 3D model selection problem.

In traditional hydrogeological conceptual model selections criteria, matching the observation data is one of the most important criterion\cite{Carrera1993201validation,Rojas2008conceptual,refsgaard2011strategy,PIROT2015influence}.  Knowing that the hydraulic conductivity has a strong correlation with the hydraulic head and through observing the changing of the hydraulic head in a transient state model with channelized hydraulic conductivity, we find the pattern of the changing of the hydraulic head is quite sensitive to the hydraulic conductivity. The head dropping at two adjacent time steps is faster where there are channels with bigger hydraulic conductivity which indicates that the feature of the changing of the hydraulic head can reflect the pattern of the hydraulic conductivity. Therefore, learning the features from the hydraulic head time series from groups of realizations generated from their corresponding TI can help us determine which TI is more applicable to the study area.

Recurrent neural networks(RNNs) have a strong capability for dealing with sequential data and have been widely applied to various real-world tasks due to its promising results, e.g., speech recognition, machine translation, time series prediction, etc. \cite{graves2013speech,sutskever2014sequence,fernandez2007application}. \citeA{chen2019gated} combined Gated Recurrent Unit(GRU) and kernel principal component analysis(KPCA) to predict the remaining useful life at a very complex system where conventional methods can't performed well.
\citeA{kumar2004river} found that RNNs had better results compared to Artificial Neural Network(ANN) in river flow forecasting.
\citeA{zhang2018developing} used Long Short-term Memory(LSTM) with water diversion, evaporation, precipitation, temperature, and time data to forecast groundwater table depth in agricultural areas. 
Inspired by the property of RNNs and their applications, we applied different RNNs, i.e., standard RNN, GRU and LSTM, to extract features from observation data of the hydraulic head at different time steps. Then a Multilayer Perceptron(MLP) was adopted to map from the extracted features to a vector that can represent predicted probability of each TI. To our best knowledge, this is the first attempt to apply RNNs to choose a proper TI by observation data of the hydraulic head.

In this paper, a new TI selection method based on RNNs is proposed for MPS algorithms when there are various TI available. Instead of running thousands of iterations as MCMC based methods, this new approach only need one time forward modeling to get the observation data, and the RNN training part is also computational efficient. Two TIs with channelized geometrical structures were applied in a hydrogeological model to test the performance of this new method. The result shows that the selection model can obtain a 97.63\% accuracy by using GRU in the synthetic hydrogeological case.

\section{Methodology}
\begin{figure}
\noindent\includegraphics[height=20pc]{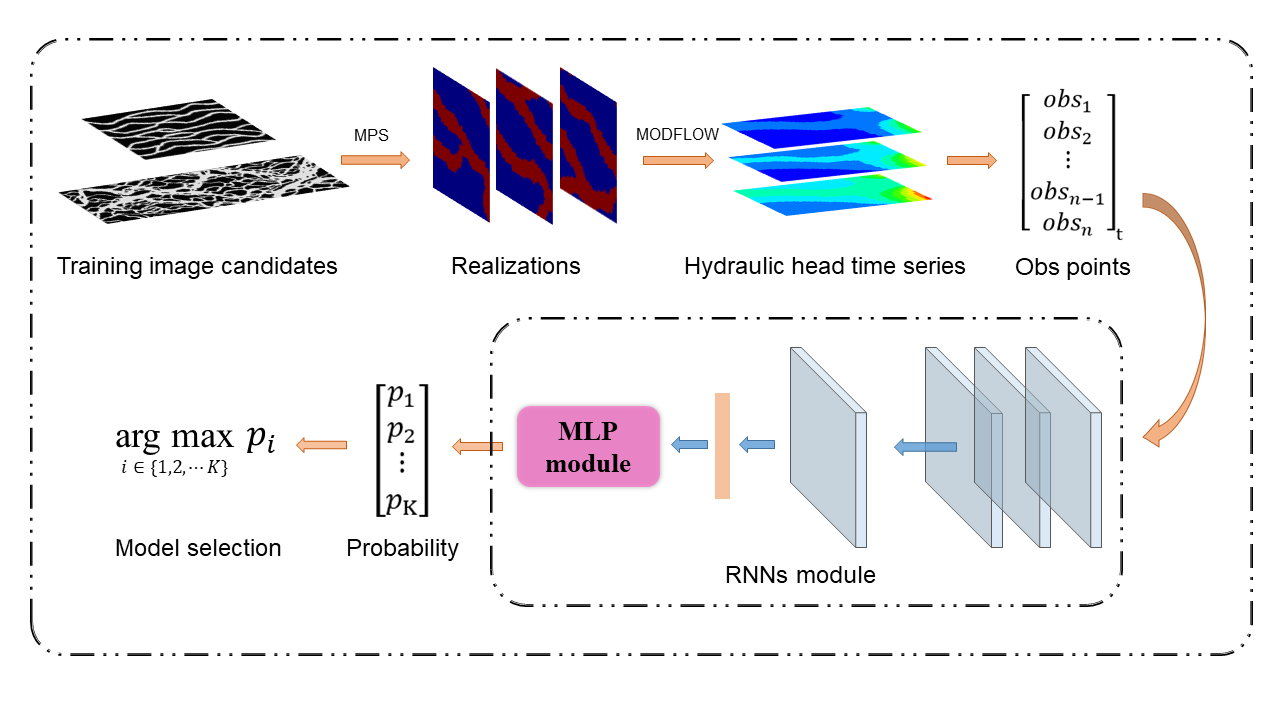}
\centering
\caption{Flowchart of TI selection method. The first step is to choose the TI candidates and then apply the MPS algorithm to generate realizations. After that, feed the MODFLOW with each of the realization as the hydraulic conductivity field and obtain the hydraulic head time series. Then, extract the data at the observation points and label the data based on the TI. Train the RNNs with the prepared data and the RNNs module will give a sequence of predicted probability of each TI and select the TI with the maximum probability. \label{fig:flowchart}}
\end{figure}

The TI selection steps can be summarized as follows: first, TI candidates selection, several TIs need to be selected or generated from field observation data to be evaluated, then, generates corresponding realizations through a MPS algorithm from each TI. After that, apply the groundwater flow equation through MODFLOW for each realization to obtain the hydraulic head time series. Label the time series respect to the corresponding TI. Next, train RNNs to extract the features of the hydraulic head time series and make the prediction decisions. After the RNN training done, enter the observations of hydraulic head into the trained model and it will select a most possible TI. Figure~\ref{fig:flowchart} shows the flowchart.

\subsection{Groundwater flow equation}

The time series were generated using the MODFLOW-2000\cite{harbaugh2000modflow} with the governing equation as follows: 
\begin{equation}
 \frac{\partial }{\partial x}\big(K_x\frac{\partial h}{\partial x}\big)+\frac{\partial }{\partial y}\big(K_y\frac{\partial h}{\partial y}\big)+\frac{\partial }{\partial z}\big(K_z\frac{\partial h}{\partial z}\big)-W^* =S_s\frac{\partial h}{\partial t}
\end{equation}
Where $K$ is the hydraulic conductivity, $h$ is the hydraulic head, $S_s$ is the specific storage and t is time. $W^*$ is the source and sink.

\subsection{Recurrent neural networks}
\begin{figure}
\centering
\noindent\includegraphics[height=15pc]{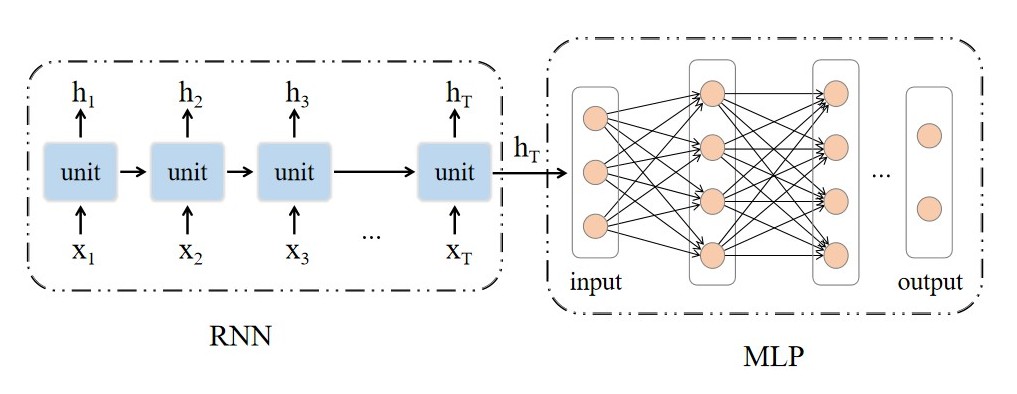}
\caption{Structure diagram of RNNs. $x_t$ denotes the observed data of the hydraulic head at time step $t$, $h_t$ denotes the hidden state at time step $t$. \label{fig:diagram}}
\end{figure}

Knowing that RNNs are efficient architectures to deal with sequential data, we applied three different RNN models to test the TI selection performance.
As Figure~\ref{fig:diagram} shows, the whole neural network is divided into two parts, RNN and MLP. First, RNN extracts features from the observation data of the hydraulic head. The extracted features are vector representations named hidden state. Then, MLP takes the hidden state at the last time step as input and maps the hidden state into another vector which can be considered as predicted probability of each TI.

The idea of RNNs is to repeatedly perform a shared unit and to update the hidden state along time steps. The shared unit is a chunk of neural network that usually takes the previous hidden state and the data at the current time step as the input, and outputs the current hidden state. All data at different time steps share the same repeating unit and a chain allows information to be passed along time steps, as is shown in picture \ref{fig:diagram}. The standard RNN, GRU and LSTM differ in the design of the repeating unit.

\subsubsection{Standard RNN}
In standard RNN, the repeating unit has a simple structure such as a single tanh layer, a linear layer with tanh activation function. Given input sequence $\textbf{x}=(x_1,x_2,\cdots,x_T)$, the hidden state is updated as the following equation iteratively from t=1 to T:
\begin{equation}
    h_t=f(W_{xh}x_t+W_{hh}h_{t-1}+b_h)
\end{equation}
where f is a non-linear activation function, $h_t$ is the current hidden state at time t, $x_t$ is the current input at time t, $h_{t-1}$ is the previous hidden state at time t-1,  $W_{xh}$ and $W_{hh}$ are repeating weight matrices, and $b_h$ denotes bias vector.

However, it has been observed by \cite{bengio1994learning} that RNN suffers from long-term dependency problems, i.e., when it comes to long sequence input, the gradients tend to vanish so that the information extracted from early observation data will be forgotten in RNN.

\begin{figure}
\noindent\includegraphics[height=14pc]{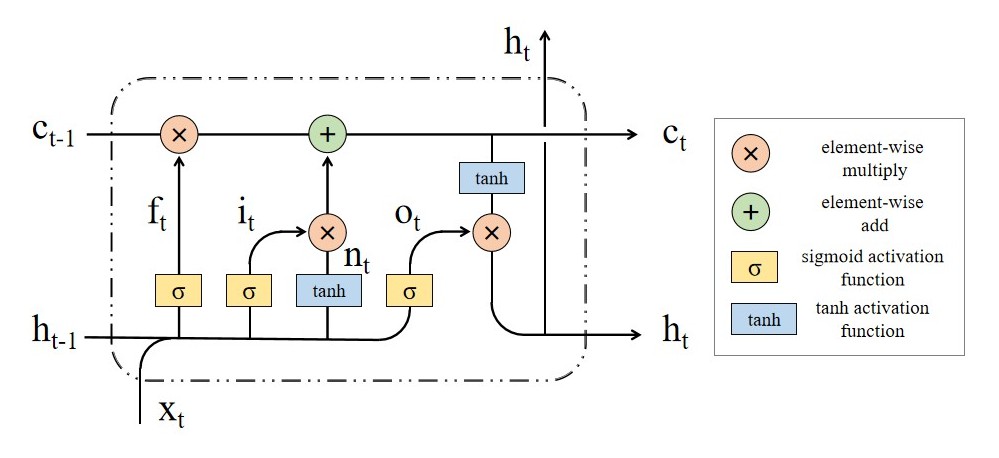}
\centering
\caption{Frame of LSTM. \label{fig:LSTM}}
\end{figure}

\begin{figure}
\noindent\includegraphics[height=15pc]{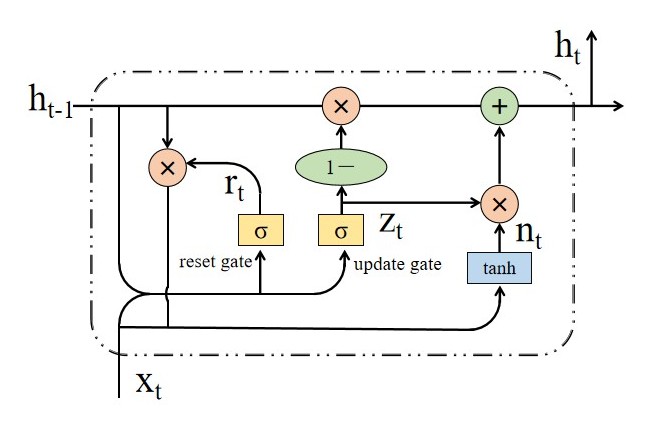}
\centering
\caption{Frame of GRU. \label{fig:GRU}}
\end{figure}

\subsubsection{Long short term memory (LSTM)}
LSTM was initially proposed by \citeA{hochreiter1997long}. LSTM unit adopts gating mechanism to help RNN avoid long-term dependency problem. The structure of LSTM unit is shown in Figure~\ref{fig:LSTM}. 

LSTM unit consists of a cell state, a hidden state and three different gates. Each gate is actually a sigmoid layer that takes the previous hidden state and current data as input. A cell state is introduced in LSTM to store historical information from previous sequential data. The extent to which the previous cell state needs to be forgotten is regulated by a forget gate. 
\begin{equation}
    f_t = \sigma(W_fh_{t-1}+U_fx_t+b_f)
\end{equation}

The new information from current input and the previous hidden state is learned by a tanh layer. Then an input gate decides the extent to which the new information needs to be added into the cell state.
\begin{eqnarray}
  n_t=tanh(W_nh_{t-1}+U_nx_t+b_n)\\
  i_t=\sigma(W_ih_{t-1}+U_ix_t+b_i)
\end{eqnarray}

Next, the cell state is updated by partially forgetting historical information and adding new information.
\begin{equation}
    c_t=f_t \odot c_{t-1}+i_t \odot n_t
\end{equation}

Lastly, the updated cell state is used to reproduce the current hidden state by a tanh layer. Then an output gate decides the extent to which the current hidden state needs to be retained.
\begin{eqnarray}
o_t=\sigma(W_oh_{t-1}+U_ox_t+b_o)\\
h_t=o_t \odot tanh(c_t)
\end{eqnarray}

In the formula above, $h_t,c_t,x_t$ are the hidden state, cell state and input at time t, and $f_t,i_t,o_t$ are the forget, input and output gates, respectively. $n_t$ is the new information. \textbf{W},\textbf{U} denote weight matrices and \textbf{b} denote bias vectors. $\sigma$ is the sigmoid function, and $\odot$ is the Hadamard product.

\subsubsection{Gated recurrent unit (GRU)}
GRU was proposed by \citeA{chung2014empirical} on statistical machine translation. Figure~\ref{fig:GRU} shows the structure of GRU.
Similar to the LSTM unit, GRU also adopts gating mechanism but without having the cell state. Instead, it mixed with cell state and hidden state. Another difference is that the forget gate and input gate are combined together into a single layer and there are only two gates in GRU, reset and update gates.

The reset gate decides how much the previous hidden state is retained to produce new information.
\begin{equation}
  r_t=\sigma(W_rh_{t-1}+U_rx_t+b_r)
\end{equation}

The new information is computed by
\begin{equation}
  n_t=tanh(W_n(r_t\odot h_{t-1})+U_nx_t+b_n)
\end{equation}

The hidden state was updated by a weighted average between the previous hidden state and the new information, and the update gate decides the weight coefficient.
\begin{eqnarray}
  z_t=\sigma(W_zh_{t-1}+U_zx_t+b_z)\\
  h_t=(1-z_t)\odot n_t+z_t\odot h_{t-1}
\end{eqnarray}

In the above formula, $n_t$ is the new information and $r_t,z_t$ are the reset and update gates, respectively. \textbf{W},\textbf{U} denote weight matrices and \textbf{b} denote bias vectors. $\sigma$ is the sigmoid function, and $\odot$ is the Hadamard product.

\subsection {Statistical interpretation}
In hydrological conceptual model selections, various studies have applied Bayesian statistics to do help make the decision\cite{Hsu2008bayesian,schonigernine2014,BRUNETTI2017127}.This new proposed neural network approach can also be interpreted in a statistical perspective. In Bayesian theory, the posterior probability, $p(Y|X)$, of a model $Y$ given data $X$ plays a crucial role in model selection. By Bayes' theory,
\begin{equation}
    p(Y|X)=\frac{p(X|Y)p(Y)}{p(X)}
\end{equation}
where $X$ denotes observations of hydraulic head, $Y$ is TI candidates, $ Y \in \{1,2,\cdots,K\}$ , $K$ is the number of TI candidates. As the term $p(X)$ is a normalizing constant of posterior distribution of $Y$, it could be neglected and the posterior probability is proportional to the likelihood function multiplied by the prior probability:
\begin{equation}
    p(Y|X) \propto p(X|Y)p(Y)
\end{equation}
The term $p(X|Y)$ named Bayesian model evidence(BME) or marginal likelihood quantifies the likelihood of observed data integrated over TI's realizations space:
\begin{equation}
    p(X|Y)=\int p(X|Y,\mu)p(\mu|Y)\mathrm{d}\mu
\end{equation}
where $\mu$ denotes realizations of each TI, $p(\mu|Y)$ represents the probability that the realization $\mu$ is generated by a given TI $Y$, the term $p(X|Y,\mu)$ represents the probability that observed data $X$ is generated by a given realization $\mu$. It's generally assumed to be uncorrelated and Gaussian distributed with constant standard deviation $\sigma$ for simplicity:
\begin{equation}
    p(X|Y,\mu)=\left( \sqrt{2\pi\sigma^2} \right)^{-n} exp \left[-\frac{1}{2}\sum_{i=1}^n \left( \frac{x_i-\mathcal{F}_i(\mu)}{\sigma}\right)^2\right] 
\end{equation}
where $\mathrm{F}(\mu)$ is the generated time series of hydraulic head given realization $\mu$ using MODFLOW with the groundwater equation, $n$ is the total number of observation data of hydraulic head. 
Since the formula 15 is hard to compute, it can be approximated by Monte Carlo method \cite{hammersley1960monte}:
\begin{equation}
    p(X|Y)=\frac{1}{N}\sum_{i=1}^Np(X|Y,\mu_i)
\end{equation}
where $\mu$ denotes the realizations of $Y$ generated through a MPS algorithm, $N$ is the number of realizations. Other modified approaches based on Monte Carlo integration such as path sampling, power posteriors, and stepping-stone sampling were proposed to compute the marginal likelihood $p(X|Y)$ in equation 15\cite{gelman1998simulating,friel2008marginal,xie2011improving}. All of them are under Bayesian frame, i.e., compute marginal likelihood first then use formula 14 to compare posterior probability of each TI. TI with large posterior probability is preferred statistically.

Our approach uses a different strategy. Instead of computing marginal likelihood first, we use neural network with parameters $\theta$ to approximate the posterior probability directly. We use a MPS algorithm to generate realizations for each TI candidates $Y$. Then, we apply the MODFLOW to produce the hydraulic head sequence $\textbf{x}$ for each realization. The neural network learn from the already matched $(\textbf{x},y)$ pairs. It takes the hydraulic head $\textbf{x}$ as an input and output a real vector $\textbf{p}$. 
$\textbf{p}=(p_1,p_2,\cdots,p_K)$ and $p_i = p(Y=i|X)$.
$p(Y=i|X)$ represents the probability of a TI $i$ given hydraulic head sequence $X$ and $\sum_{i=1}^K p(Y=i|X)=1$. Obviously,
\begin{equation}
    p(Y=i|X)=\prod_{i=1}^K p(Y=i|X)^{I_{\{i\}}(Y)}
\end{equation}
where $I_{\{i\}}(Y)$ is the indicator function, s.t.,

\begin{equation}
I_{\{i\}}(Y)=
\begin{cases}
1, & Y=i\\
0, & otherwise 
\end{cases}
\end{equation}

Actually, the training of the neural network is to find a proper group of parameters $\theta$, s.t., 
\begin{equation}
    \hat{\theta}=\mathop{\arg\max}_{\theta} \log(p(Y|X,\theta)) = \mathop{\arg\min}_{\theta} -\log(p(Y|X,\theta)) = \mathop{\arg\min}_{\theta} Loss
\end{equation}

We choose cross entropy error as loss function, which is equal to the negative logarithm of the posterior probability. Specifically,
\begin{equation}
    Loss=-\frac{1}{N}\sum_{(\textbf{x},y)\in S } \sum_{i=1}^K
    I_{\{i\}}(y) \log p(y=i|\textbf{x})
\end{equation}
where $S$ is the dataset, $N$ is the dataset size, $(\textbf{x},y)$ are the matched data pairs of hydraulic head sequence and TI labels in dataset $S$.
In prediction, the trained neural network takes the observation data of hydraulic head as an input and output the TI predicted vector $\textbf{p}$. The selected TI is:
\begin{equation}
    \hat{Y}=\mathop{\arg\max}_{i \in \{ 1,2,\dots,K\}}p_i
\end{equation}

\subsection{Networks training}
All networks were trained in Pytorch, an open source machine learning library supporting strong graphics processing units(GPU) acceleration \cite{paszke2019pytorch}. We used the Adam algorithm to search a proper group of parameters to minimize the cross entropy loss in formula 20. Adam is one of the gradient-based optimization algorithms that are based on calculating the gradients of loss function with respect to the weights of network. It can be considered as the combination of AdaGrad and RMSProp and has a faster convergence rate compared to other gradient-based algorithms, e.g., SGD, AdaGrad and RMSProp \cite{Kingma2014Adam}. 
Gradients involved in Adam can be computed by back-propagation through time(BPTT), an extension of back-propagation(BP) to deal with RNNs \cite{rumelhart1986learning,hecht1992theory,werbos1990backpropagation}. Gradients in Pytorch can be computed by automatic differentiation technique \cite{paszke2017automatic}. 

The dataset was divided into training set(80\%) and testing set(20\%). We trained the network on the training set and tested its performance on the testing set. The $L_2$ regularization and dropout techniques were used to prevent the overfitting problem,i.e., models have much better performance on the training set than the testing set \cite{hawkins2004problem}.
$L_2$ regularization is a popular weight decay approach that drives the weights of the network to be closer to 0. It has been observed that a weight decay can improve the generalization of neural network \cite{krogh1992simple}.
The key point of $L_2$ regularization is adding a penalty term to the loss function. The main idea of the dropout technique is to randomly drop out nodes of neural network during train process and it could improve neural network performance by preventing units from co-adapting  \cite{hinton2012improving,srivastava2014dropout}.
Batch Normalization(BN) and learning rate exponential decay techniques were also adopted to accelerate the convergence. BN is a widely-used technique that makes the training of neural network faster and more stable. Despite its promising property, the mechanism of BN remains under discuss. A widely accepted explanation is that BN can reduce the internal covariate shift problem, i.e., the change in the distributions of layers' inputs makes upper layers hard to learn \cite{ioffe2015batch}. By fixing the distribution of layers' inputs, BN can improve training speed. \citeA{santurkar2018does} proposed that BN improves training speed by smoothing the optimization landscape rather than reducing internal covariate shift.
\citeA{cooijmans2016recurrent} found that it's beneficial to apply BN to the hidden state transition in RNNs.

Since model performance is influenced by the hyper-parameters of the network, We adopted random search strategy to search a proper group of hyper-parameters. The hyper-parameters are chosen the best one from at least 50 different groups of hyper-parameters for each model. The hyper-parameters cover learning rate, dropout probability, regularization coefficient, dimension of hidden state, batch size, period and multiplicative factor of learning rate decay. We used ray.tune, a unified platform for model selection that provides hyper-parameters searching algorithms, to help search hyper-parameters \cite{liaw2018tune}.
Part of hyper-parameters setting are shown in Table~\ref{tab:hyper-parameters}. 

\begin{table}
 \caption{Hyper-parameters setting. Lambda denotes regularization coefficient, hidden size denotes dimension of hidden state, gamma denotes multiplicative factor of learning rate decay and step size denotes period of learning rate decay. \label{tab:hyper-parameters}}
 \centering
 \begin{tabular}{cccccccc}
 \hline
  model  & learning rate & dropout & lambda  & hidden size & batch size & gamma & step size \\
 \hline
  RNN & 0.0001 & 6.09 $\times 10^{-2}$ & 1.97 $\times 10^{-3}$ & 128 & 64 & 0.5 & 40 \\
  GRU & 0.0001 & 1.37 $\times 10^{-2}$ & 4.28 $\times 10^{-4}$ & 128 & 128 & 0.2 & 40 \\ 
  LSTM & 0.0001 & 7.04 $\times 10^{-3}$ & 1.45 $\times 10^{-4}$ & 128 & 128 & 0.5 & 40 \\
 \hline
\end{tabular}
 \end{table}

\section{Data}

The data set is constitute with two steps. The first step is generating realizations from each TI by MPS algorithms. The second step is generating the hydraulic head time series by MODFLOW for each realization.

\subsection{Realizations of TI candidates}

\begin{figure}
\centering
\noindent\includegraphics[height=23pc]{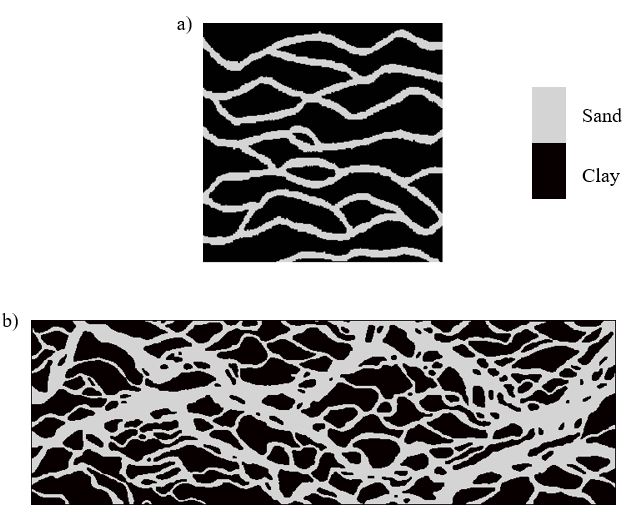}
\caption{TI candidates. \label{fig:TI} }
\end{figure}

\begin{figure}
\centering
\noindent\includegraphics[height=18pc]{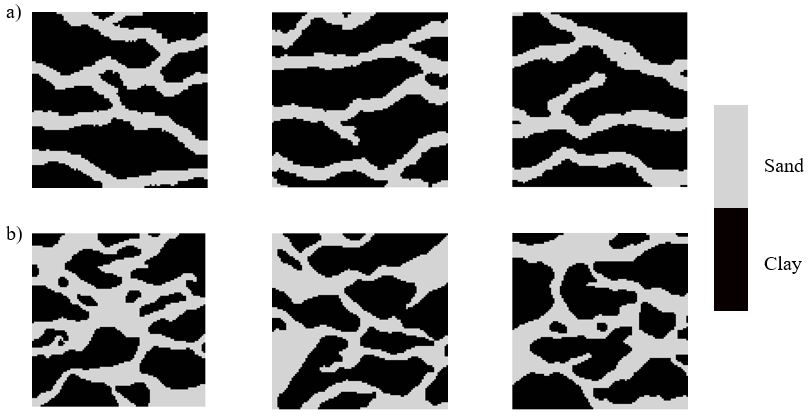}

\caption{Realizations of corresponding TIs. Row a is the corresponding realizations of TI a, and row b is the corresponding realizations of TI b.  \label{fig:relization} }
\end{figure}

Two TI candidates of a potential fluvial originated aquifer are selected to generate realizations to obtain the corresponding hydraulic head time series.TI a (Figure~\ref{fig:TI} (a)) is a classic channelized TI obtained from \cite{strebelle2002conditional}, TI b (Figure~\ref{fig:TI} (b)) is obtained from the TI Library and represents the Bangladesh delta. In both TIs, the channel is assumed fill with sand and the background is filled with clay. Figure~\ref{fig:relization} shows the corresponding realizations of each TI generated by the Quick Sampling\cite{mathieu2020qs}. The input parameters of the Quick Sampling is shown in Table~\ref{tab:MPS}.

Notice here, we chose two structurally similar TIs in order to create more challenges for the selection task. In TI a, the channels are in same width while in TI b, the channels' width varies from place to place. However, both of them show a strong connectivity and curvilinear structures.

\begin{table}
\caption{Input Parameters of the QS.\label{tab:MPS}}
\begin{center}
\begin{tabular}{ll}
\hline
   Training Image a size & 250 $\times$ 250 \\
   Training Image b size & 768 $\times$ 243 \\
   Simulation size&    100 $\times$ 100 \\
   Search Radius & 50 \\
   Number of best candidates &1.2 \\
   Data type & Categorical\\
\hline
\end{tabular}
\end{center}
\end{table}

\subsection{Time series of hydraulic head}
\begin{figure}
\noindent\includegraphics[width=1.2\linewidth]{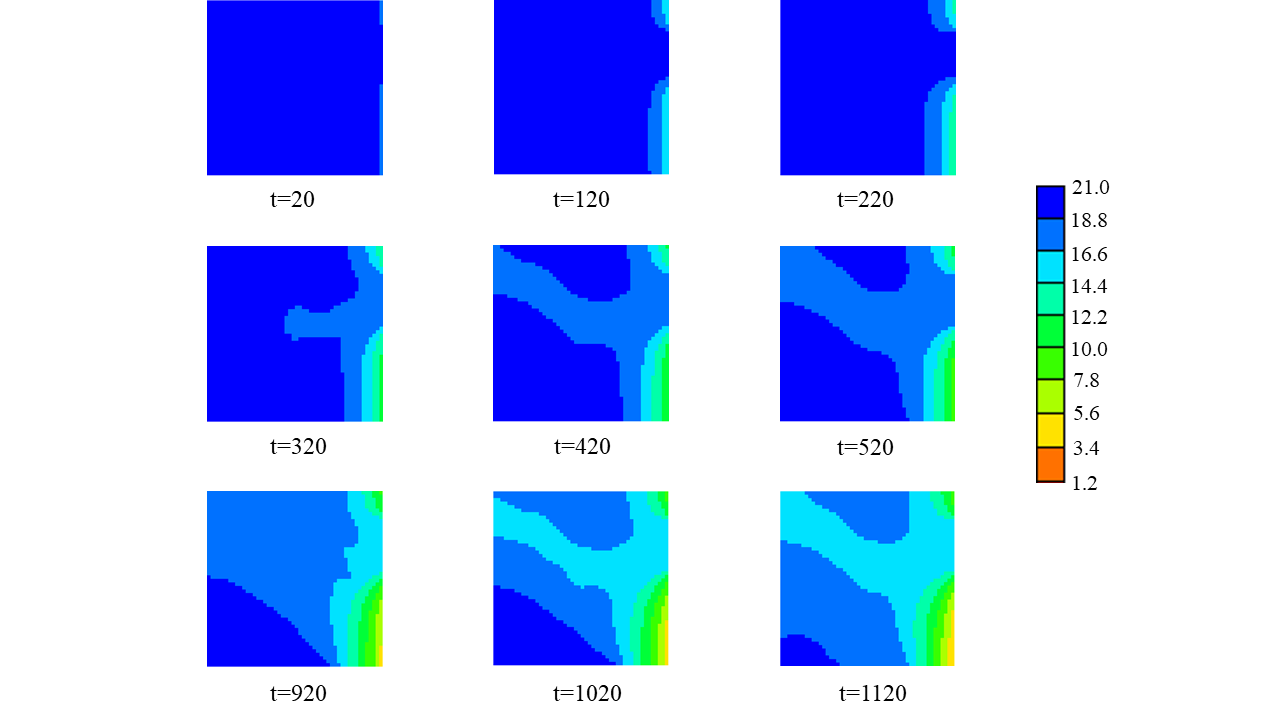}
\caption{Selected hydraulic head time series. \label{fig:head} }
\end{figure}

A transient flow model was applied to run the test. A period of 2000 days were divided into 100 time steps. The aquifer has 100 $\times$ 100 $\times$ 1 cells and each cell has a size of 1 m $\times$ 1m $\times$ 1m. The boundary conditions of the aquifer were simplified as no flow boundary on the North, South and East sides. 100 pumping wells were set on the West boundary with pumping rate $q= -0.1$ $m^3 /day$ to get the time series of the hydraulic head. The hydraulic conductivity(K) for sand was set to $K_{sand}= 3.5$ $m /day$, and clay was set to $K_{clay}= 0.004$ $m /day$. Different numbers observation wells were selected to get the observed hydraulic head data. Table 1 shows the aquifer setting parameters. Figure~\ref{fig:head} is an illustration of hydraulic head time series at different time steps for one realization. As the figure shows, the pattern of the the changing of the hydraulic head can reflect the channel structure in a certain degree. 

\begin{table}
\caption{Setup of the groundwater flow model.\label{tab:GW}}
\begin{center}
\begin{tabular}{ll}
\hline
   Model size&    100 $\times$ 100 \\
   Grid size &  1 m $\times$ 1 m \\
   Aquifer thickness and top elevation & 10 m \\
   Simulation time & 2000 days \\
   Number of periods & 1 \\
   Number of time steps  & 100 \\
   Aquifer storage coefficient & 0.003\\
   Initial head & 20 m \\
\hline
\end{tabular}
\end{center}
\end{table}

\section{Results}

\subsection{Evaluation metrics }
We used three metrics to evaluate the performance of all models.
\begin{enumerate}
    \item \textbf{Accuracy} The percentage of correct prediction for model selection.
    \item \textbf{AUC} Area Under the ROC Curve(AUC) which equals to the probability that a classifier will rank a randomly chosen positive instance higher than a randomly chosen negative one \cite{fawcett2006introduction}. A higher AUC value indicates a better performance.
    \item \textbf{Cross entropy loss} Since the cross entropy loss is the objective function that all models attempted to minimize, we used it as a straightforward metric.
\end{enumerate}

The accuracy and AUC value on testing set are two main metrics.

\subsection{Performance of models}
\begin{figure}
\noindent\includegraphics[width=\textwidth]{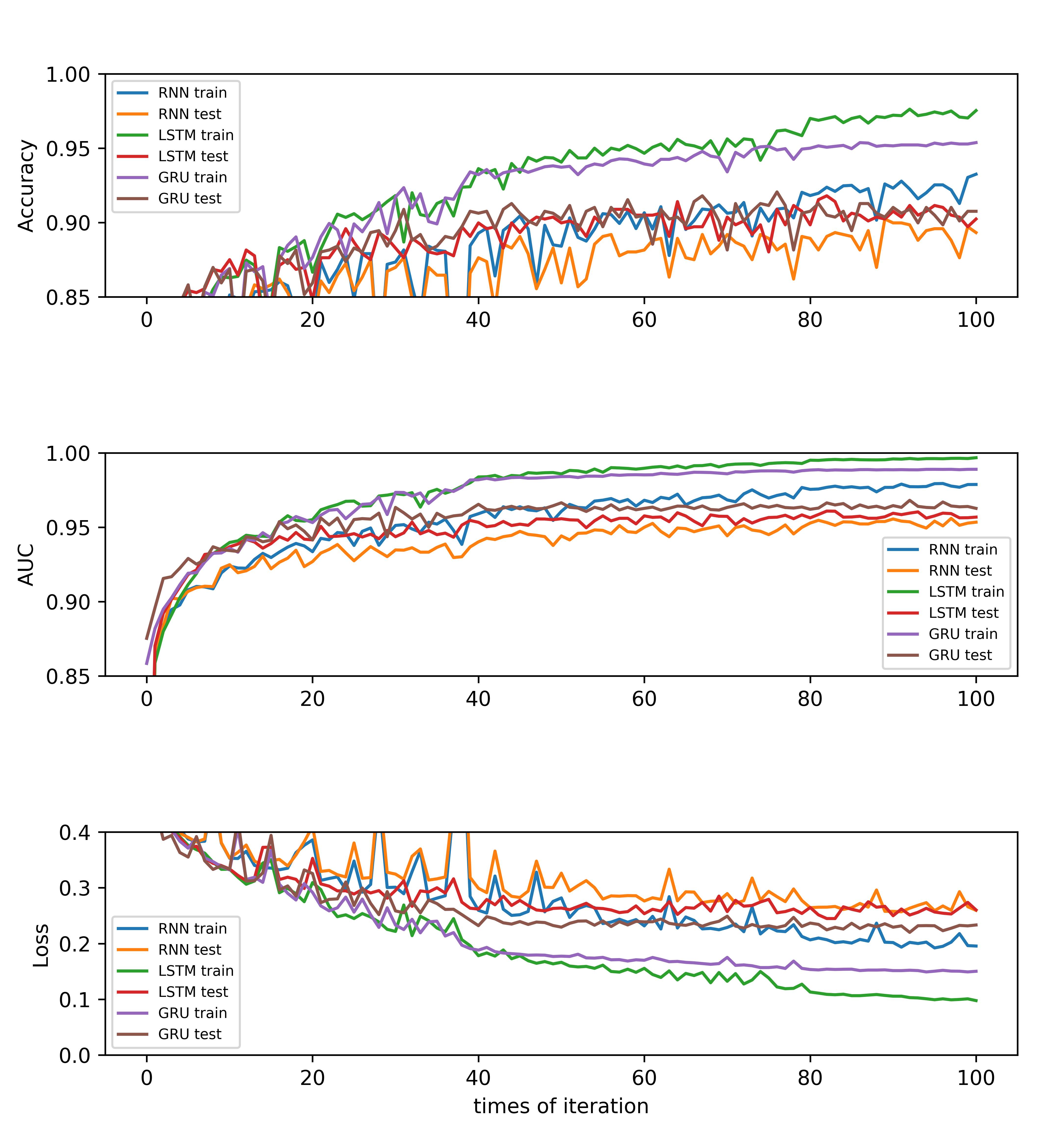}
\caption{Model comparison on data set with 49 observations and sequence length of 100. We trained each model with 50 different groups of hyper-parameters and chose the best one to present respectively. \label{fig:model}}
\end{figure}

We compared the performance of three different RNNs on the data set with 49 observation wells and 100 time steps of the hydraulic head. The results are shown in Figure~\ref{fig:model} and Table~\ref{tab:model}. All of the three different RNNs acquire an over 90$\%$ accuracy on the testing set, which indicates that this method has a high accuracy in TI selection given with the observation data. In addition to that, according to Figure~\ref{fig:model}, GRU and LSTM achieve a better performance than the standard RNN. The reason is that standard RNN suffers from long-term dependency problems, while GRU and LSTM can alleviate them with the gating mechanism. The overfitting problem still exists although we tried L2 regulation and dropout techniques with different combinations of regularization coefficient and dropout probability. That's possibly because dataset size of 4000 is relatively small for the selection task that overfitting problem is hard to avoid. 
Comparing to LSTM, GRU performs slightly better on testing data with 1.19$\%$ rise on accuracy, 0.0089 rise on AUC and 0.0326 decline on loss. It seems that GRU is a more efficient model to extract information from the observation data of hydraulic head but the impact of different RNNs architectures is small.

 \begin{table}
 \caption{Model comparison on data set with 49 observation wells and 100 time steps. All the results were computed by each model with parameters that made each test accuracy highest. \label{tab:model}}
 \centering
 \begin{tabular}{ccccccc}
 \hline
  model  & train accuracy & test accuracy & train AUC & test AUC & train loss & test loss \\
 \hline
  RNN & 0.9372 & 0.9023 & 0.9831 & 0.9592 & 0.1729 & 0.2440 \\
  GRU & 0.9559 & \textbf{0.9219} & 0.9897 & \textbf{0.9698} & 0.1458 & \textbf{0.2122} \\ 
  LSTM & \textbf{0.9700} & 0.9180 & \textbf{0.9954} & 0.9609 & \textbf{0.1088} & 0.2448 \\
 \hline
\end{tabular}
 \end{table}

\subsection{Influence of number of observation wells}
The number of observation wells of hydraulic head $N$ is an important factor which has a directly impact to the TI selection accuracy . In this new approach, $N$ is also the input dimension of RNNs. We compared GRU performances on different data sets with $N$=9, 49, 400, 900 and 1600 respectively. Since the appropriate hyper-parameters may vary with the change of number of observation wells, we tried 50 different groups of hyper-parameters for each N and choose the best one, i.e., the group of hyper-parameters that have highest accuracy on testing set.

\begin{figure}
\centering
\noindent\includegraphics[width=\textwidth]{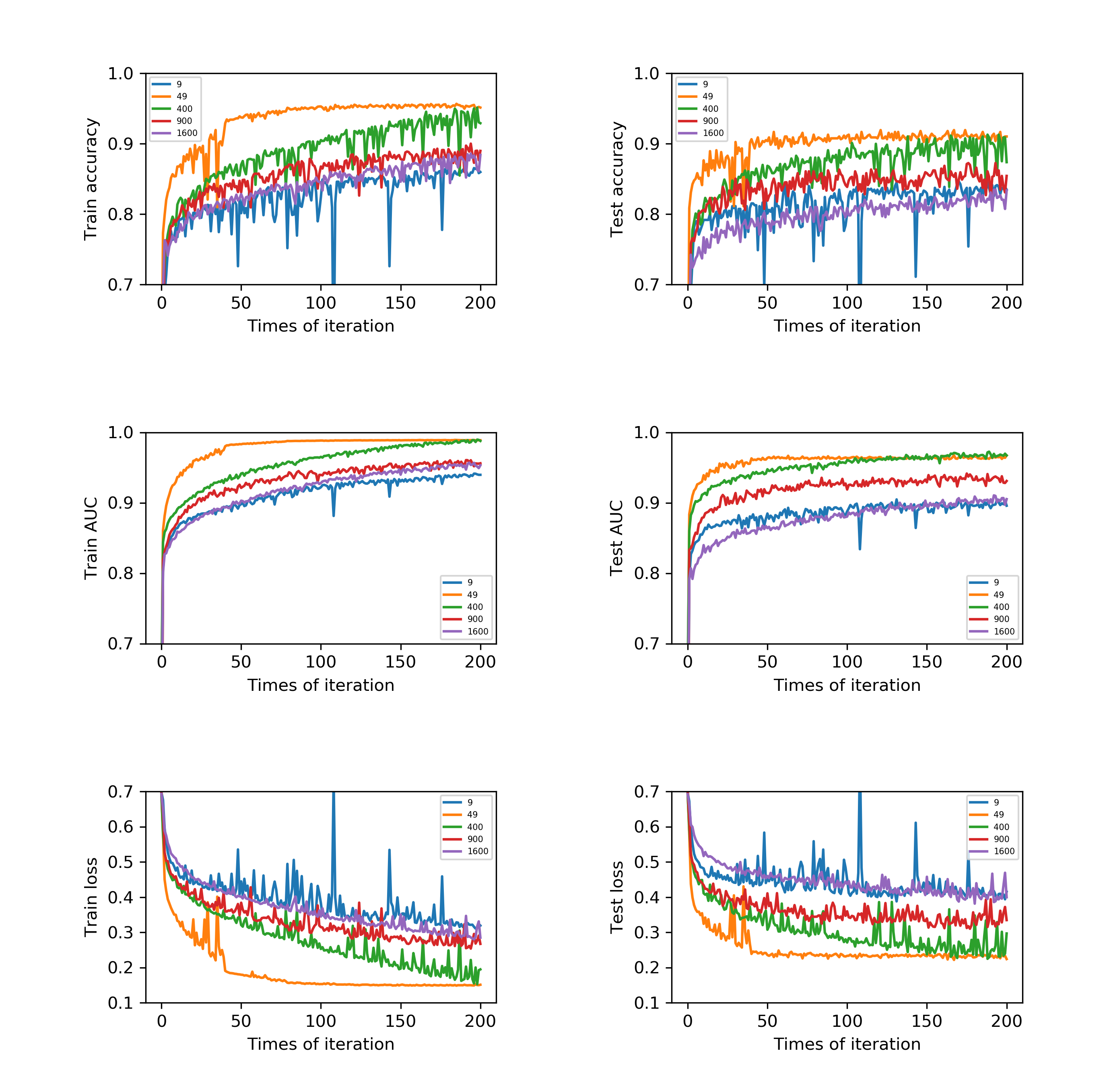}

\caption{GRU performances on data set with sequence length of 100 and different observation wells of hydraulic head sequence.
\label{fig:obs} }
\end{figure}

The results are shown at Table~\ref{tab:observation} and Figure~\ref{fig:obs}. As $N$ increases from 9 to 400, the test accuracy increases from 85.55\% to 92.19\%, and test AUC increases from 0.9073 to 0.9705. It intuitively makes sense because the more observation wells of hydraulic head, the more information is provided to select TI. However, as $N$ increase from 400 to 1600, the tendency is adverse, i.e., the test accuracy decreases to under 89\%, and test AUC decreases to under 0.95. An possible explanation is that a large input dimension may make the network easier to fall into local extreme points. Nevertheless, in practice applications, observations of hydraulic head may mot be abundant enough to cause this issue due to the economic cost.

 \begin{table}
 \caption{GRU performances on data sets with different numbers of observation wells. 50 different groups of hyper-parameters are tested for each number of observation wells and the one with highest accuracy on the testing set are selected. Again, all the lines were presented with parameters that made each test accuracy highest. \label{tab:observation}}
 \centering
 \begin{tabular}{ccccccc}
 \hline
  observation wells & train accuracy & test accuracy & train AUC & test AUC & train loss & test loss \\
 \hline
  9 & 0.8678 & 0.8555 & 0.9446 & 0.9073 & 0.3025 & 0.3878 \\
  49 & 0.9559 & \textbf{0.9219} & 0.9897 & 0.9698 & 0.1458 & \textbf{0.2122} \\ 
  400 & \textbf{0.9700} & \textbf{0.9219} & \textbf{0.9985} & \textbf{0.9705} & \textbf{0.0966} & 0.2440 \\
  900 & 0.9034 & 0.8841 & 0.9664 & 0.9412 & 0.2348 & 0.3082 \\
  1600 & 0.9631 & 0.8880 & 0.9953 & 0.9421 & 0.1056 & 0.3549 \\
 \hline 
\end{tabular}
 \end{table}

\subsection{Influence of sequence length}
In order to verify the impact of sequence length (time steps) $L$ of hydraulic head on TI selection, we tested different sequence length varies from 20, 50 to 100 time steps and kept other conditions unchanged. Notice here, all different length of time series was selected from the end of the sequence. Again, since the change of sequence length may influence the choice of hyper-parameters, we tried 50 different groups of hyper-parameters for each sequence length and choose the one with highest accuracy on testing data.

\begin{figure}
\centering
\noindent\includegraphics[width=\textwidth]{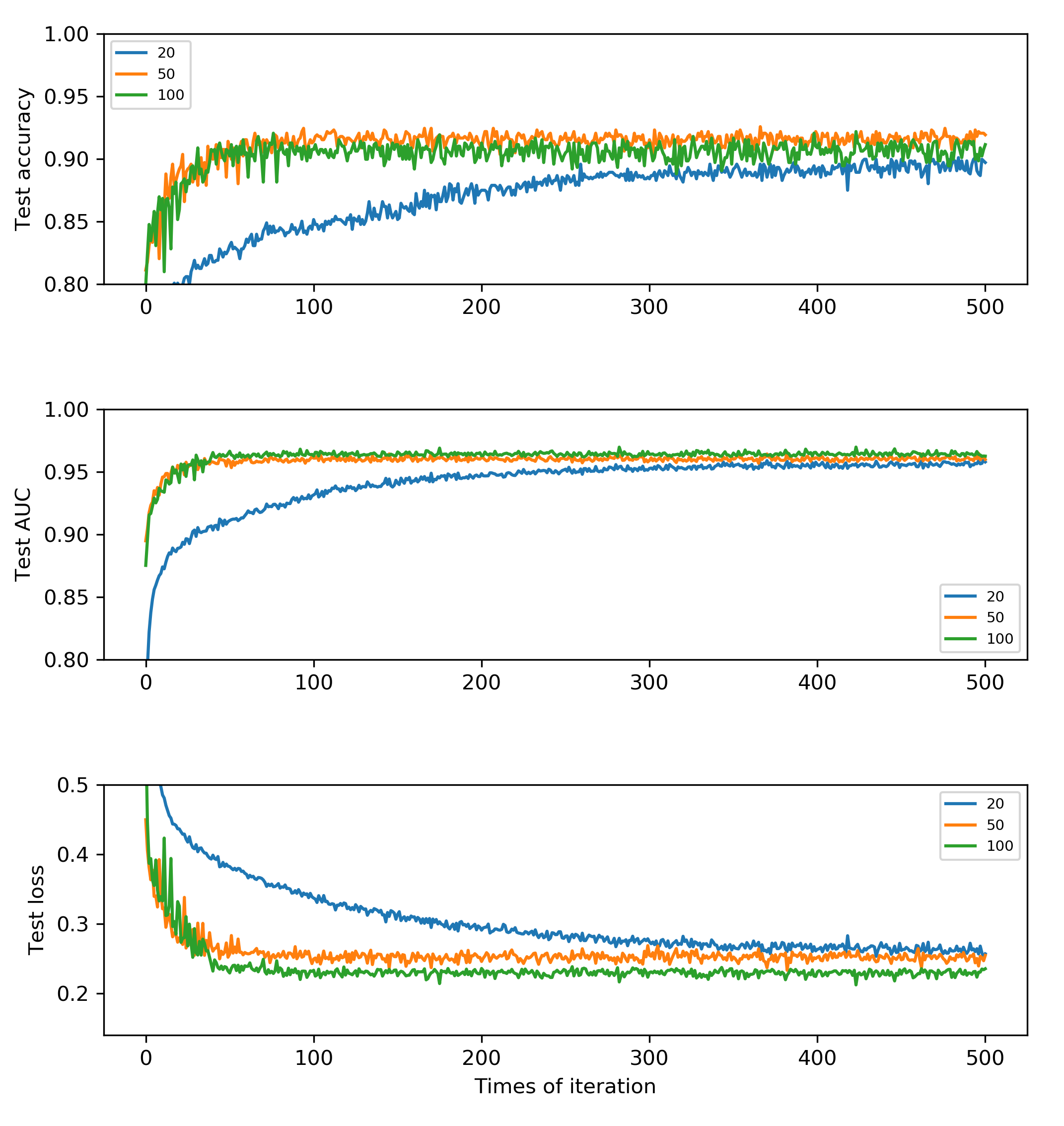}

\caption{GRU performances on testing set with 49 observation wells and different length of hydraulic head sequence.
\label{fig:steps} }
\end{figure}

Table~\ref{tab:length} and Figure~\ref{fig:steps} show the GRU performances on data set with $L=$ 20, 50 and 100. We can see that GRU performances on testing set with $L=$ 50 and 100 are very close, i.e., as $L$ increases from 50 to 100, test accuracy decreases 0.39\%, test AUC increasing 0.0085 and test loss decreasing 0.0365. GRU performance on testing set with $L=$ 20 is poorer than $L=$ 50 and 100, but the difference is small. The experimental results indicate that the sequence length of hydraulic head has little impact on TI selection.

\begin{table}
 \caption{GRU performances on data sets with different sequence length of hydraulic head.  \label{tab:length}}
 \centering
 \begin{tabular}{ccccccc}
 \hline
  sequence length & train accuracy & test accuracy & train AUC & test AUC & train loss & test loss \\
 \hline
  20 & 0.9300 & 0.9010 & 0.9795 & 0.9579 & 0.1943 & 0.2555 \\
  50 & \textbf{0.9819} & \textbf{0.9258} & \textbf{0.9982} & 0.9613 & \textbf{0.0940} & 0.2487 \\ 
  100 & 0.9559 & 0.9219 & 0.9897 & \textbf{0.9698} & 0.1458 & \textbf{0.2122} \\
 \hline 
\end{tabular}
 \end{table}

\subsection{Influence of observation noise}
\begin{figure}
\centering
\noindent\includegraphics[width=\textwidth]{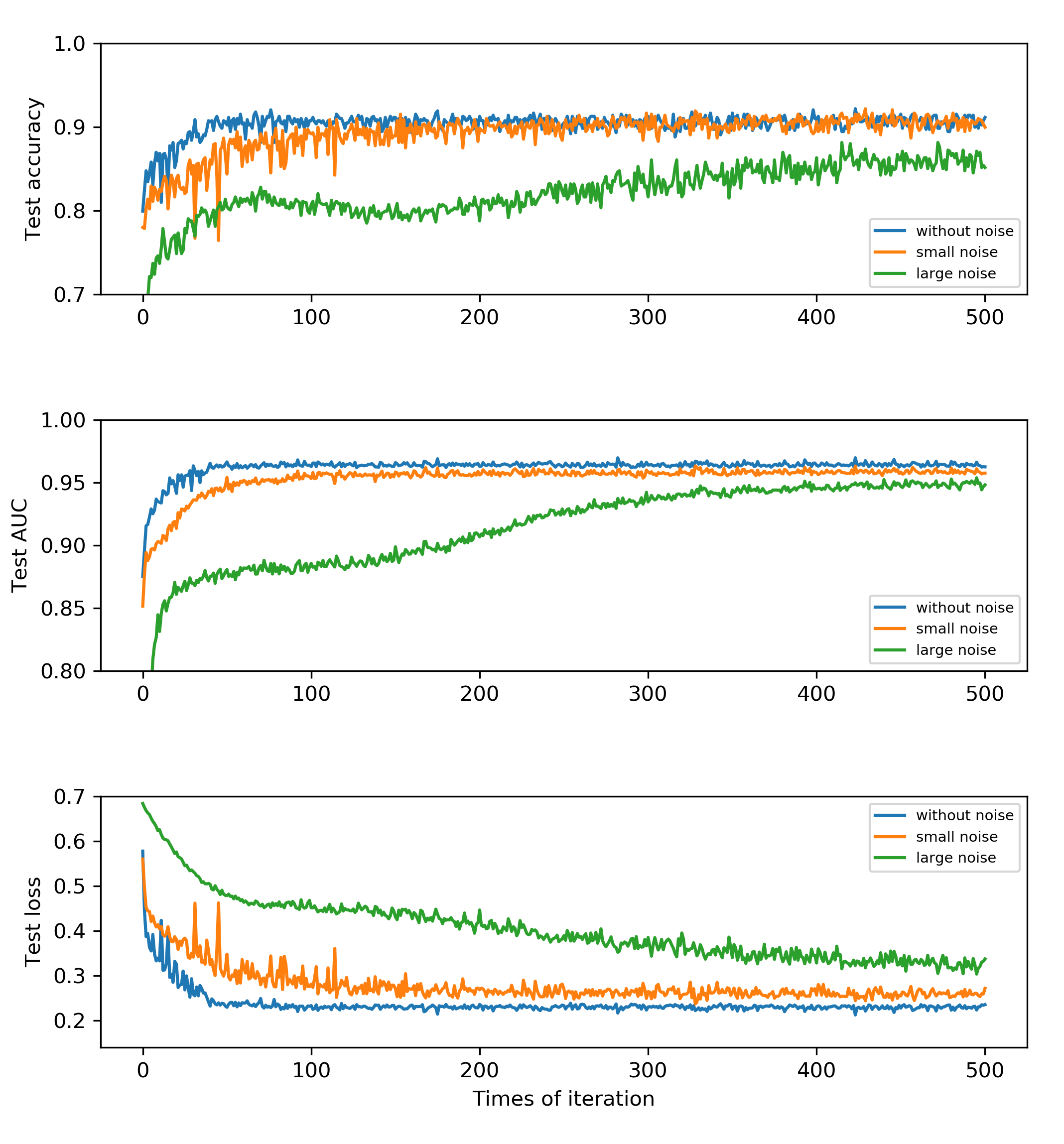}

\caption{GRU performances on testing set with 49 observation wells, 100 time steps and different noise setting. Small noise denoted uncorrelated Gaussian noise with a small variance $\sigma^2=1$ and a mean $\mu=0$, while large noise denotes uncorrelated Gaussian noise with a small variance $\sigma^2=9$ and a mean $\mu=0$
\label{fig:noise} }
\end{figure}

Considering that there are usually errors in collecting hydraulic head observation data in real applications, we added two different Gaussian noise to hydraulic head sequences and compared their performances to the case without adding noise. One is an uncorrelated Gaussian noise with a small variance $\sigma^2=1$ and a mean $\mu=0$, the other is an uncorrelated Gaussian noise with a large variance $\sigma^2=9$ and a mean $\mu=0$. The results are shown in Table~\ref{tab:noise} and Figure~\ref{fig:noise}. It's clear that adding a small Gaussian noise to observations of hydraulic head doesn't degrade the GRU performances on both training and testing set. However, when adding a large noise to observations of hydraulic head, we can see a visible degradation of GRU performance, i.e., 7.4\% decline on train accuracy, 4.04\% decline on test accuracy, 0.0385 decline on train AUC, 0.0244 decline on test AUC, 0.1552 rise on train loss and 0.1031 rise on test loss. 

It can also been seen from Table~\ref{tab:noise} that the GRU performances on training set and testing set are very close when adding a large noise to observations of hydraulic head. It seems to indicate that adding a random noise can alleviate the overfitting problem for the neural network to some extent and thus the degradation of GRU performance on testing set is significantly lighter than on training set.

\begin{table}
 \caption{GRU performances on data set with 49 observation wells, 100 time steps and different noise setting. Small noise denoted uncorrelated Gaussian noise with a small variance $\sigma^2=1$ and a mean $\mu=0$, while large noise denotes uncorrelated Gaussian noise with a large variance $\sigma^2=9$ and a mean $\mu=0$ \label{tab:noise}}
 \centering
 \begin{tabular}{ccccccc}
 \hline
  noise & train accuracy & test accuracy & train AUC & test AUC & train loss & test loss \\
 \hline
  without noise & \textbf{0.9559} & \textbf{0.9219} & \textbf{0.9897} & \textbf{0.9698} & \textbf{0.1458} & \textbf{0.2122} \\
  small noise & 0.9406 & \textbf{0.9219} & 0.9865 & 0.9599 & 0.1728 & 0.2439 \\ 
  large noise & 0.8819 & 0.8815 & 0.9512 & 0.9454 & 0.3010 & 0.3153 \\
 \hline 
\end{tabular}
 \end{table}

\subsection{Influence of dataset size}

\begin{figure}
\noindent\includegraphics[width=\textwidth]{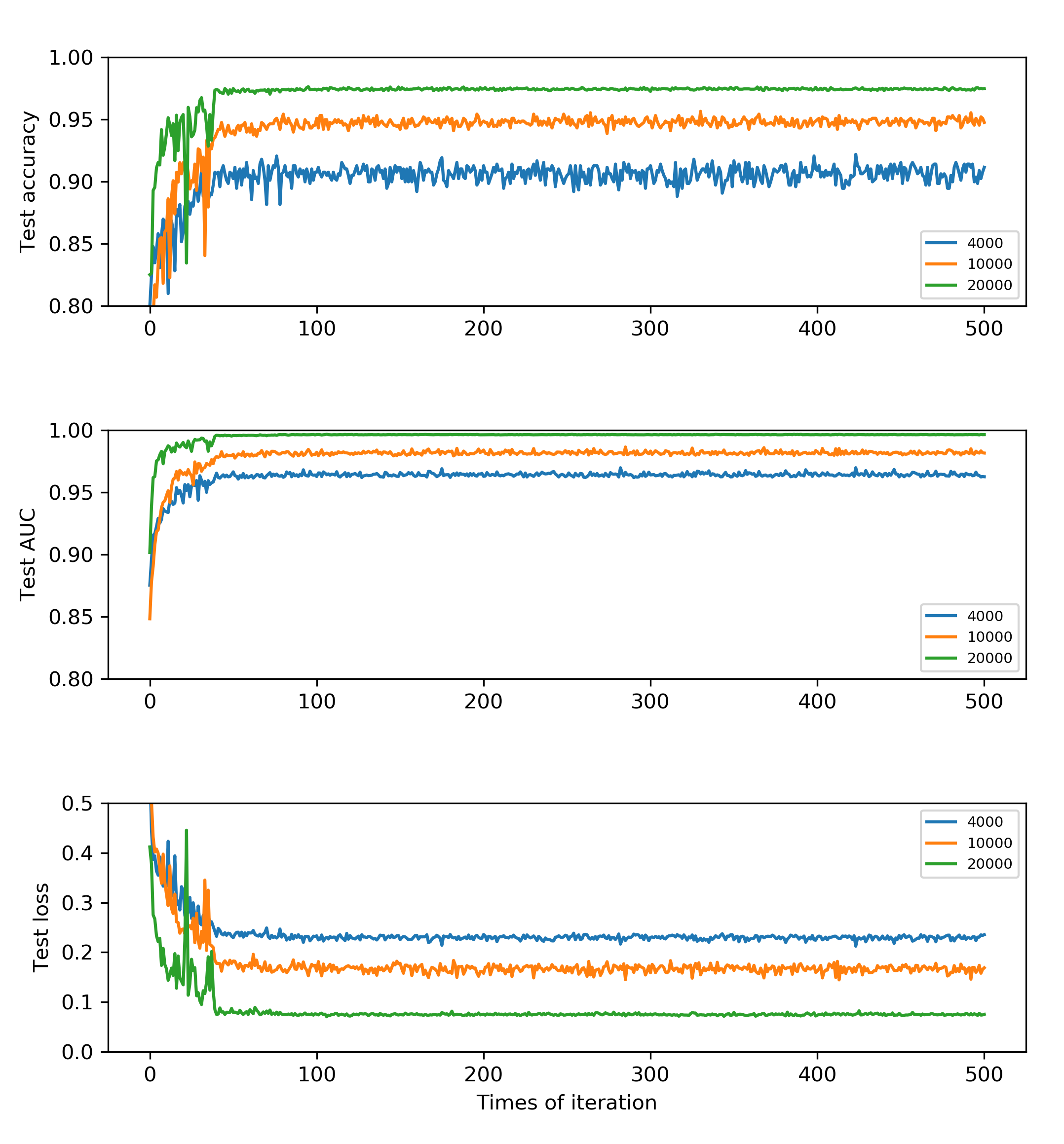}
\caption{GRU performances on data set of different size. Other conditions such as sequence length, observation number and hyper-parameters are kept unchanged. \label{fig:samples}}
\end{figure}

Three different sizes(4000,10000,20000) of dataset are tested to see the model performance. The hyper-parameters  maintain unchanged for all sizes of dataset. As Table~\ref{tab:samples} and Figure~\ref{fig:samples} show, GRU performances on both training set and testing set are greatly improved when the dataset size increases. As the dataset size increses from 4000 to 20000, The test accuracy is increasing from 92.19\% to 97.63\%, the test AUC is increasing from 0.9698 to 0.9963, and test loss is decreasing from 0.2122 to 0.0755.
In addition, as Figure~\ref{fig:samples} shows, the tendencies of the curves are pretty consistent. They have the inflection point almost at the same position.  

\begin{table}
 \caption{GRU performances on data set of different size. Other conditions such as sequence length, observation number and hyper-parameters are kept unchanged. \label{tab:samples}}
 \centering
 \begin{tabular}{ccccccc}
 \hline
  Dataset Size & train accuracy & test accuracy & train AUC & test AUC & train loss & test loss \\
 \hline
  4000 & 0.9559 & 0.9219 & 0.9897 & 0.9698 & 0.1458 & 0.2122 \\
  10000 & 0.9922 & 0.9565 & 0.9995 & 0.9831 & 0.0425 & 0.1547 \\ 
  20000 & \textbf{0.9982} & \textbf{0.9763} & \textbf{1.000} & \textbf{0.9963} & \textbf{0.0121} & \textbf{0.0755} \\
 \hline 
\end{tabular}
 \end{table}

\section{Conclusions}

In this paper, we proposed a robust and straight forward TI selection model using RNNs to help researchers to determine an applicable TI in a given context for their study areas. We have presented the work flow of the new method as well as the detailed information about neural network settings. 

Three different RNNs architectures are tested to select a proper TI among two channelized TI candidates. Results show that GRU performs slightly better than LSTM, while LSTM performs better than standard RNN on both prediction accuracy and AUC value. Number of observation points of hydraulic head plays a significant role in our method. Enough observations points can improve the accuracy of TI selection while too much points may cause an adverse effect. 50 time steps of the hydraulic head are sufficient enough for the GRU to extract the features and do the TI selection job. Adding a small noise to observations of hydraulic head doesn't degrade model performance. Even in the case with large noise the prediction accuracy is still acceptable. This is a promising property because there may be errors between observed data and the true value in practice but it has little effect on this proposed method. Increasing the size of dataset can greatly improve the accuracy of TI selection in our method and 16000 training data has a satisfactory performance when facing with the overfit issue. Since there is no iteration needed in the forward simulation in this new framework, increasing dataset size won't significantly increase the computational cost. 

The concept of this method is to capture the relationship between the observation data and the hard data, and based on that relationship, to make selection choice. An example has been illustrated via a groundwater model. The result shows that the selection accuracy can reach to 97.63\%. TI selection plays a significance role in the whole workflow of modeling a complex subsurface system. This method can be considered in the foremost step in a MPS based workflow. Besides, this selection model can also be applied to other subsurface models such as oil reservoir models, CO\textsubscript{2} sequestration models.

\acknowledgments
Training images in this paper are downloaded from http://www.trainingimages.org/training-images-library.html. The Quick Sampling algorithm can be obtained from https://github.com/GAIA-UNIL/G2S.


%
%

\bibliography{ TIRNN }

%
%
%
%
%

\end{document}